# Some remarks on the mathematical structure of the multiverse

Alan McKenzie

*Retired: formally of University Hospitals Bristol NHS Foundation Trust, Bristol UK*

**Abstract**

The Copenhagen interpretation of quantum entanglement experiments is at best incomplete, since the intermediate state induced by collapse of the wave function apparently depends upon the inertial rest frame in which the experiment is observed. While Everett's Many Worlds Interpretation (MWI) avoids the issue of wave function collapse, it, too, is a casualty of the special theory of relativity.  This requires all events in the universe, past, present and future, to be unique, as in the block-universe picture, which rules out Everett-style branching.  The benefits of MWI may be retained, however, by postulating a multiverse of discrete, parallel, block universes which are identical to each other up to certain points in the MWI "trunk" before they diverge according to the MWI branching.  The quantum probability of an event then emerges from the number of parallel universes in which the event happens divided by the total number of universes.  This means that the total number of such universes is finite.  Such a picture is more easily envisaged by thinking of it as a purely mathematical structure, as in Tegmark's Mathematical Universe Hypothesis. However, while Tegmark wished to avoid contamination from Gödelian self-referential knots, not only does such contamination appear to be inevitable, it brings an unexpected benefit.  The mathematical hierarchy required by Gödel's enigmatic footnote 48a leads to an explanation for a unitary evolution of deterministic quantum rules across the multiverse while accounting for quantum uncertainty within an individual universe.  Other aspects of this structure, called here the Plexus, are discussed, including awareness of existence and other questions raised by the hypothesis.

1.      **Collapse of the wave function**

Nearly a century after Niels Bohr and Werner Heisenberg separately wrote down their understanding of the elements of quantum mechanics, controversy over the Copenhagen interpretation is still very much alive: a quick *arXiv* search reveals a remarkable number of papers still debating the issue.  Although there is no definitive list of Copenhagen principles, a consensus evolved from the 1950s which generally included effective collapse of the wave function with overtones of non-locality.  It is the collapse aspect that raises difficulty.

Consider, for instance, the spins of a pair of electrons, entangled in a singlet state, measured respectively by Alice and Bob. The electrons travel in opposite directions at equal speeds from a source that is closer to Alice than to Bob (see Figure 1). They each orientate their spin filter at a randomly chosen angle "at the last moment" (so that it is too late for information about the respective angles of inclination to be swapped between them before the electrons arrive at the filters: their measurements are separated by a space-like interval so that the experiment is non-local). We suppose that Alice chooses to orientate her filter vertically at 0° and that Bob chooses an angle of 120° for his filter.

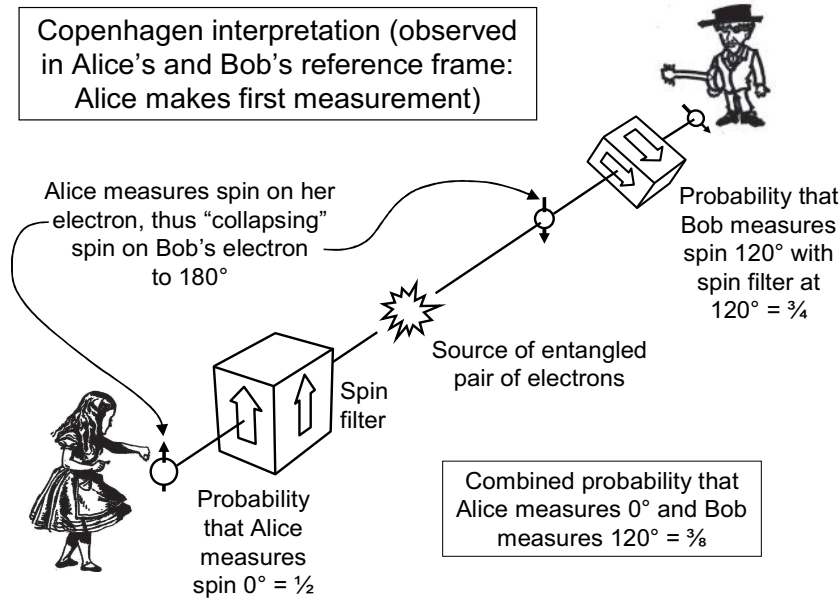

**Figure 1**: In this reference frame, the source of the pair of electrons is closer to Alice than to Bob and so she makes the first measurement.

Alice makes the first measurement, the source of electrons being closer to her. The standard Copenhagen interpretation of the experiment is that there is a probability of ½ that she will find her electron has a 0° spin, and that the act of measurement collapses the wave function, setting the spin of the other electron, heading towards Bob, at 180°. When this electron reaches Bob's filter, inclined at 120°, the probability that Bob will measure a spin of 120° is $\cos^2[(180°-120°)/2] = ¾$, so that the combined probability that Alice measures 0° and Bob measures 120° is ⅜.

Since the interval between Alice's and Bob's measurements is space-like – non-local – it is always possible to find a frame of reference in which the order of measurements is reversed. So, in a reference frame travelling sufficiently fast from Alice towards Bob, it is Bob, and not Alice, who makes the first measurement (see Figure 2). The Copenhagen view of this is that there is a probability of ½ that Bob will find his electron spin orientated at 120° and that the act of measurement collapses Alice's electron (still en route to her) at 300°. When this electron reaches Alice's spin filter, the probability that Alice will measure 0° is $\cos^2[(360°-300°)/2] = ¾$, and, again, the combined probability that Alice measures 0° and Bob measures 120° is ⅜.

So the Copenhagen process can be used to calculate the combined probability of the two measurements on the entangled system, regardless of the frame of reference used to make the calculation. However, collapse produces two pictures that are quite different: in the first frame, we have an electron with a spin of 180° heading towards Bob and in the second we have an electron with spin 300° heading towards Alice. Clearly, both pictures cannot be real. Since all inertial frames are regarded equally in special relativity, there cannot be a favoured frame and so neither picture can be real.

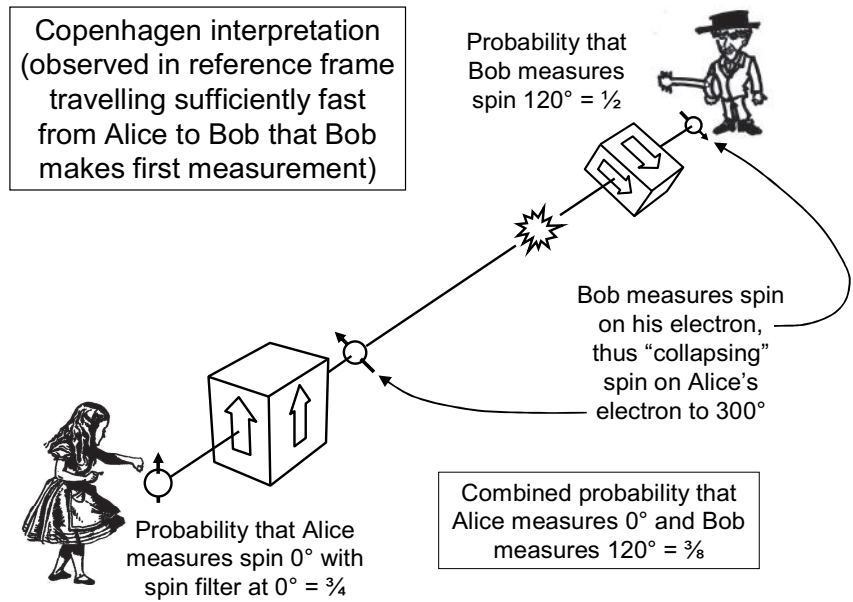

**Figure 2**: In this frame of reference, it is Bob who makes the first measurement.

## 2. Many Worlds Interpretation

Difficulties such as this led to a quest for alternative views, among which Everett's Many Worlds Interpretation (MWI) [1] has remained a respectable contender for decades [2], [3]. This has the attraction that there is no need for the wave function to collapse; instead, we consider the unitary evolution of the wave function of the whole universe in accordance with the Schrödinger equation, forever branching into orthogonal, non-interfering states. In many of these branches a version of oneself exists, similar or identical to the versions in other branches, and quantum uncertainty is manifest, for instance, by an individual version not knowing which branch it inhabits. Many workers have discussed measurement, decoherence and how probabilities might be quantified in MWI [4], [5], [6], [7], [8], [9], [10], [11], [12], [13], but there is no general consensus.

Figure 3 uses the above entanglement experiment to illustrate some of the features of MWI. Alice and Bob may measure their electron spin to be up or down, represented respectively by Au, Ad, Bu and Bd. (Here, "Bu" means Bob measuring a 120° spin when his spin filter is orientated at 120° and "Bd" means him finding a 300° spin.)

The upper-left "tree" shows the experiment from Alice's point of view. There is a time when she has made her measurement but news of Bob's outcome, carried at the speed of light, has not yet reached her, and so the two possibilities are labelled "AuB?" and "AdB?".

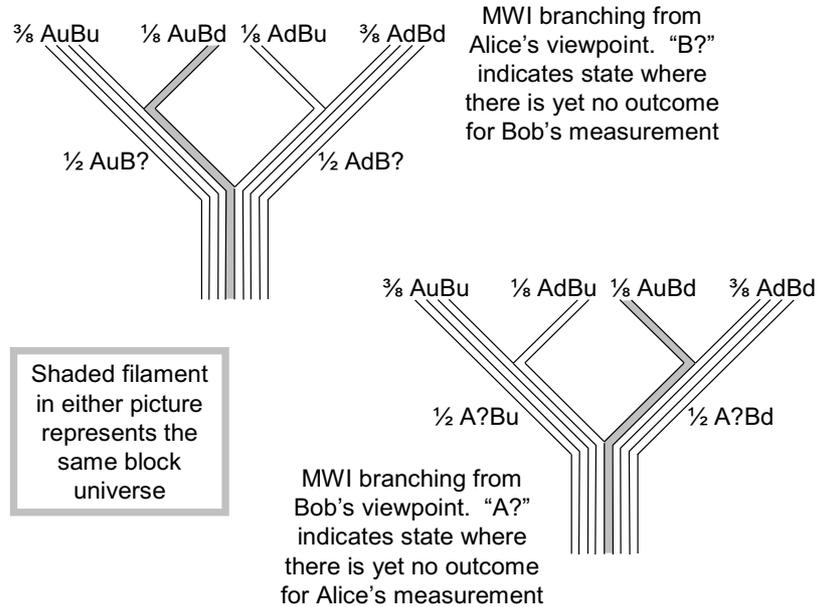

**Figure 3**: The topology of branching in MWI depends upon the viewer, but the final outcomes are the same: a probability of ⅜ that both Bob and Alice will find an up-spin, a probability of ⅜ that they will both find a down-spin, a probability of ⅛ that Alice will find an up-spin while Bob finds a down-spin, and a probability of ⅛ that Alice will find a down-spin while Bob finds an up-spin.

All four possible outcomes are shown at the top of the tree, and the outcomes are each labelled with a fraction to indicate its relative probability. Each fraction may be regarded as the absolute square of the amplitude for that particular outcome. Notice that the thicknesses of the trunk and branches of the tree have been drawn proportionally to the relative probabilities. The tree has been sectioned into equal-sized filaments to make this easier to see. The tree to the lower-right shows the same experiment from Bob's viewpoint. While there is a difference in the topology of the branching between the two viewpoints, the final outcomes are the same as may be seen, for instance, by comparing the two highlighted filaments in the two trees.

### 3. Block universe

While issues such as branch thicknesses and topologies invite debate, the branching structure of MWI provokes a deeper concern, rooted once again in Einstein's theory of special relativity. It is the block universe.

Suppose that Bob is at a distance $x$ from Alice, as measured in her reference frame, and moves at constant speed, $v$, towards her and that they each clap their hands at the same time in Alice's reference frame. These two clapping events are not simultaneous in Bob's reference frame – according to the Lorentz transformations, Alice would have to clap her hands at a later time, $vx/c^2$, for them to be simultaneous for Bob. In other words, for Alice, at the moment she first claps her hands, the time interval, $vx/c^2$, while yet in her future, is already in Bob's past. This experiment can be repeated by other observers at other places and times and in other reference frames so that, over an appropriate span of reference frames, all of Alice's future is already in the past. Since the past cannot be changed, neither can Alice's future. This is the block-universe view, where all events, past, present and future, are woven into the four-dimensional fabric of the spacetime block. It is remarkable how small an impact this astonishing conclusion has made upon general culture[1].

But now we have a problem. Our block universe is incompatible with MWI. Although we do not know the outcome of a quantum measurement that we are about to make, there are reference frames in our universe in which we have already made the measurement and observed the outcome – and so that outcome is unchangeable and therefore unique. There is no branching in a block universe.

### 4. A multiverse of parallel, block universes

However, there is a way to retain the principal advantage of MWI (*viz.*, it gets round the need for wave-function collapse) by using a topologically different structure that is nevertheless mathematically equivalent. We can take each filament of the MWI tree, such as the one highlighted in each of the two trees in Figure 3, and regard it as a block universe. So there are as many block universes as there are filaments in the tree. These parallel, stand-alone, block universes do not interact at any point in their structures. Figure 4 shows the block-universe representation of the entanglement experiment.

The relative numbers of block universes with different outcomes are determined by the absolute squares of the amplitudes for those outcomes. Notice that only one of the eight block universes has Alice detecting spin-up while Bob detects spin-down (right-most universe in the top row). This corresponds to the block universe highlighted in each of the two trees in Figure 3.

Of course, every measurement or quantum event in the universe occurs against a background of a myriad of other such events, and so the eight parallel universes in Figure 4 must be multiplied by every group of parallel universes resulting from these other events. Let us call each group of parallel universes corresponding to a particular event a "kernel". The kernel for the entanglement experiment is shown in Figure 4 as

---

[1] Although, for all his somewhat equivocal attitude to science, and, indeed, occasional disparaging remarks about Einstein himself, T S Eliot captures a key aspect of the block universe remarkably well in the opening lines of his (admittedly religious and mystical) "Four Quartets":

> Time present and time past
> Are both perhaps present in time future,
> And time future contained in time past.
> If all time is eternally present
> All time is unredeemable.

$K_{\text{AB entangled}}$ = (3AuBu + 1AuBd + 1AdBu + 3AdBd). If Alice and Bob repeat the same experiment $N$ times, then the kernel for the combined group of experiments is given by $K_{\text{Group}} = K_{\text{AB entangled}}^{\otimes N}$. In a series of such experiments in a typical block universe, the observed relative proportions of the four possible combinations of Alice and Bob finding spin-up or spin-down will more closely approach the proportions in the individual kernel as $N$ becomes very large. In other words, in the majority of universes, after many experiments, Alice and Bob will deduce that, for instance, it is three times more likely that their spins in an individual experiment will match rather than not. Similarly, the kernel for the combination of all events in a universe is the tensor product of the kernels for all of the individual events.

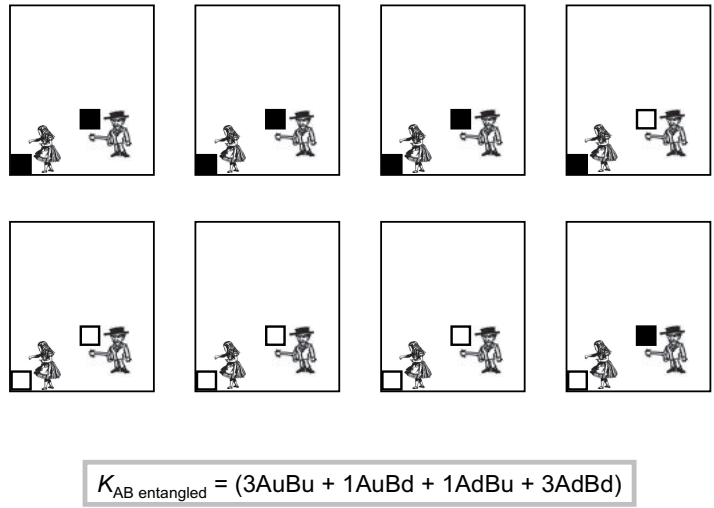

$K_{\text{AB entangled}}$ = (3AuBu + 1AuBd + 1AdBu + 3AdBd)

**Figure 4**: Parallel, block-universe representation of the experiment in Figures 1-3. A black square represents an outcome of spin-up and an empty one represents spin-down.

A property of these individual parallel, block universes is that they are very similar to each other at early times (corresponding to positions low on the tree trunk) and that differences between them become more marked at later times (corresponding to higher levels in the branching).

Implicit in this description of a multiverse of parallel, block universes is that their number is finite. An infinite number would entail the measure problem [14], involving probabilities that are ratios of infinities. This is at variance with MWI, where the number of branches at the top of the tree (and, hence, the number of filaments running right down to the trunk) may be infinite, and even uncountably so[2].

The consequence of requiring that the number of parallel universes in the multiverse be finite is that probabilities must always be rational numbers, as in the example in

---

[2] See the curious response of Hugh Everett to a question put to him by Boris Podolsky: "It looks like we would have a non-denumerable infinity of worlds". To this quite profound statement Everett simply utters a laconical "Yes" [15].

Figure 4, where the probability of both Alice and Bob measuring spin-up is ⅜. However, the thickness, or weight, of branches in MWI is determined by the Schrödinger equation, which gives rise to real, rather than rational, probabilities. For the Schrödinger equation to generate the rational numbers demanded by the parallel, block-universe model, it would need to be recast as a digital equation. (See, for instance, Gerard 't Hooft, who outlines an approach to devising a quantum theory where, rather than using real numbers for fundamental quantities like position and momentum, integers are used instead [16]. See also Ramin Zahedi, who has reformulated the field equations of the strong, electromagnetic and gravitational forces (and other equations from quantum theory) in terms of integers and matrices: he prefaces his paper with a nice review of work in the field [17].)

### 5. The multiverse is purely a mathematical structure

While MWI has too many branches to be a perfect template for the parallel, block-universe model, it nevertheless underlines the need to regard the multiverse as a purely mathematical structure. An ultimate description of the universe or the multiverse cannot depend upon components (such as fundamental particles) that are actually within the universe or multiverse it is trying to describe. In the final analysis, the language of any such description has to be independent of its subject, and elemental mathematics (*i.e.*, basic mathematical relations, freed from the symbolism used by humankind to express mathematical discoveries) is the natural candidate.

Max Tegmark is the current principal protagonist and populariser of the concept of a mathematical universe [18], [19], [20], [21]. His approach is to define a hierarchical mathematical structure containing four levels of multiverse. Our universe, along with many others, is embedded in the Level I multiverse, whilst the most general and fundamental part of the structure is Level IV. Tegmark wishes to protect his structure from any contamination from Gödelian self-referential knots in the mathematics:

> "I have long wondered whether Gödel's incompleteness theorem in some sense torpedoes the MUH [Mathematical Universe Hypothesis]" [19]

The resolution, according to Tegmark, is that only mathematical statements that are *decidable* can qualify as components of the mathematical multiverse. Tegmark calls this scheme the *Computable Universe Hypothesis*. Gödel statements – undecidable propositions – are prohibited in his scheme.

### 6. Gödel's enigmatic footnote 48a

However, quantum rules, whether digital or real, rely upon a simple, formal system of arithmetical rules (for instance, addition, in the case of superposition of states). So, from this perspective, the mathematical structure containing our universe and multiverse has to contain undecidable propositions. The resolution of these undecidable propositions is contained in Gödel's enigmatic footnote 48a:

> "The true source of the incompleteness attaching to all formal systems of mathematics, is to be found – as will be shown in Part II of this essay – in the fact that the formation of ever higher types can be continued into the transfinite…" [22]

In other words (I shall return to the "transfinite" term in a moment), in any sufficiently complex system, there are undecidable propositions that can be decided at higher levels which incorporate new axioms from which the propositions now follow either immediately or through theorems incorporating the lower axioms. Alfred Tarski came to the same understanding independently of Gödel four years later:

> "All sentences constructed according to Gödel's method possess the property that it can be established whether they are true or false on the basis of the metatheory of higher order having a correct definition of truth." [23]

To put that prosaically in terms of our own block universe, events will occur (such as the discovery that the orientation of an electron is spin-up) which cannot be predicted with certainty from anything within the block universe itself, but which follow logically from appropriate rules within a higher system. So the unitary, deterministic quantum rules may be regarded as axioms of the multiverse: from the viewpoint of the multiverse, events in specific individual universes are uniquely determined, whereas they can only be probabilistic from the perspective of any given universe.

Equally, of course, there are facts about the multiverse that cannot be deduced from within the multiverse itself. An example is the mass of the electron and the corresponding value of the Higgs field: these are apparently not provable using the quantum rules of our multiverse and yet they are clearly truths. However, in the eternal-inflation model [24], [25], our multiverse (regarded as our universe in the eternal-inflation model because quantum effects were not themselves attributed to a multiverse) is but one of an infinite multitude, in each one of which the Higgs field may take a different value.

Just as quantum effects within our own universe may be explained by additional quantum axioms at the level of our multiverse, these different values for the Higgs field may again be seen as being determined at a yet higher level where further axioms generate the process of eternal inflation and the production of a multitude of multiverses (one of which is our own).

## 7. Mathematical structure of the Plexus

In the light of this, we can no longer refer to *the multiverse* when we mean the complete mathematical structure, which must extend beyond our own multiverse. Instead, I shall use the word *Plexus* for the complete structure, capitalized in the same spirit as we capitalize the "Earth" and the "Sun". Of course, our understanding of the detailed structure of the Plexus is subject to change – for instance, a new theory may supersede eternal inflation – but the basic concept of the Plexus as a hierarchy of levels emerging from Gödelian considerations is likely to be more robust.

Figure 5 illustrates this hierarchy with the number of axioms increasing in the upwards direction. In this figure, the block universe at the lowest level refers to the whole cosmos, and is not limited to the observable universe. At this level, axioms define the geometry of the block universe, and will presumably contain the Einstein field equations (appropriately digitized as discussed above [17]). Unexplained events in the block universe are determined at a higher level by additional axioms that generate parallel block universes with events distributed according to quantum rules. At a higher level still, further axioms describe a multitude of such multiverses, accounting for different constants of nature. The ascending spiral of additional axioms terminates, according to Gödel, in the transfinite, although this permits the self-reference that Tegmark wanted to avoid. Clearly, Figure 5 is very schematic, and, for instance, the levels of the multiverse and the "multitude" could be swapped.

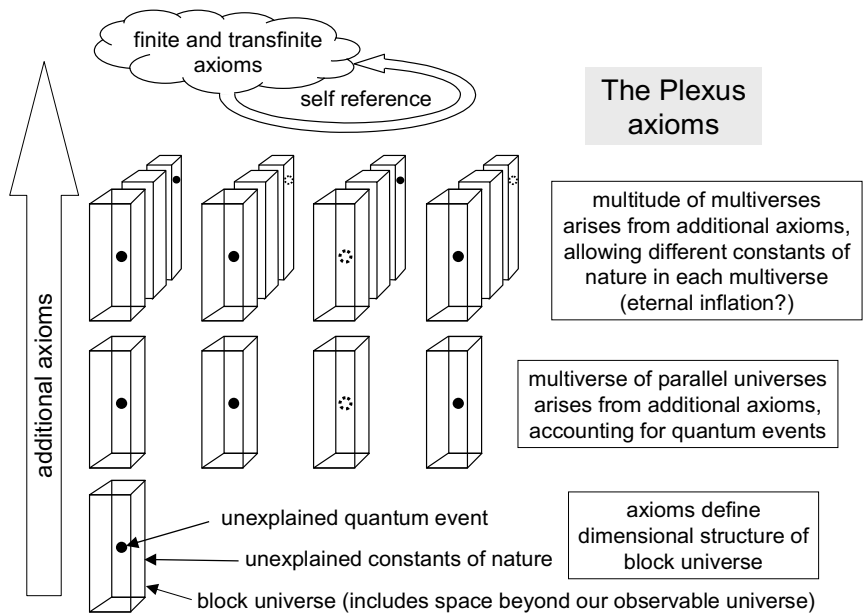

**Figure 5**: Schematic of the Plexus. Quantum events in our block universe are determined along with those in parallel block universes by additional axioms at the multiverse level. Further axioms at a yet higher level define a multitude of multiverses like our own. The ascending spiral of axioms eventually terminates in the transfinite, which permits self-reference.

## 8. I compute, therefore I am

It can be difficult to accept the idea that we are purely a mathematical structure. How could a piece of mathematics be aware of itself? To try to answer this, consider the mathematical structure in Figure 6.

While this is a relatively simple structure, we might be more comfortable thinking of it – through dimensional chauvinism on our part – as a set of matrices defined by spatial coordinates $p$ and $q$ and which are stacked in the vertical direction, $h$. The cells in the matrices are either 1 or 0, which we can think of as "live" or "dead" respectively. If we regard $h$ as the time dimension, then the matrices could be displayed sequentially on a computer screen, starting at the bottom with the initial, given state, *State*(*p*,*q*,0), and moving upwards in the direction of increasing *h* (i.e. increasing time).

If we program our computer to display this structure sequentially, then, depending upon the configuration of the initial state, we see cells being born and dying, groups of cells scuttling across the screen, bouncing off or annihilating each other and generally giving the impression of a world bustling with life. In fact, this is John Conway's Game of Life, described by Martin Gardner [26], who spells out simple rules governing whether cells live or die on the next move, depending upon the number of live cells next to them. Remarkably, it is possible to design Turing Machines within the Game of Life, a feat accomplished by Paul Rendell in 2000 [27], following it up a decade later with a Universal Turing Machine (UTM) [28].

$$-\infty \leq p, q \leq \infty$$
$$0 \leq h \leq \infty$$

$$State\,(p,q,0) = 0 \text{ or } 1$$

$$N(p,q,h) = \left\{ \sum_{a=p-1}^{p+1} \sum_{b=q-1}^{q+1} State(a,b,h) \right\} - State(p,q,h)$$

$$State\,(p,q,h+1) = \begin{cases} 1 & \text{if } N(p,q,h) \leq 3 \text{ and } (3 - State\,(p,q,h) \leq N(p,q,h)) \\ 0 & \text{otherwise} \end{cases}$$

**Figure 6**: This relatively simple mathematical structure is sufficient to describe a viable toy universe.

It is important to keep in mind that displaying and running the Game of Life sequentially, frame after frame, is only a convenience for ourselves – the computer program is isomorphic to the *static* mathematical structure in Figure 6.

Suppose that we run a Game of Life featuring two UTMs – one called UTM and the other embedded in a creature called Ant-1. This creature can, in principle, detect features of its environment, for example, by sending out groups of radar cells (called "gliders", discovered by Richard Guy in 1970) and can, through its internal UTM, build a model of its environment. Now let us program UTM to run a second Game of Life, "Game of Life 2" featuring a second Ant, called Ant-2 (see Figures 7 & 8).

We can see Ant-1 on our computer screen beside UTM. Looking closely into the UTM we can even see the individual gliders and other automata buzzing around as time progresses on the computer. However, we have not programmed a computer to show Ant-2 in the Game of Life 2, and so this is represented schematically in Figure 8 by a thought-bubble.

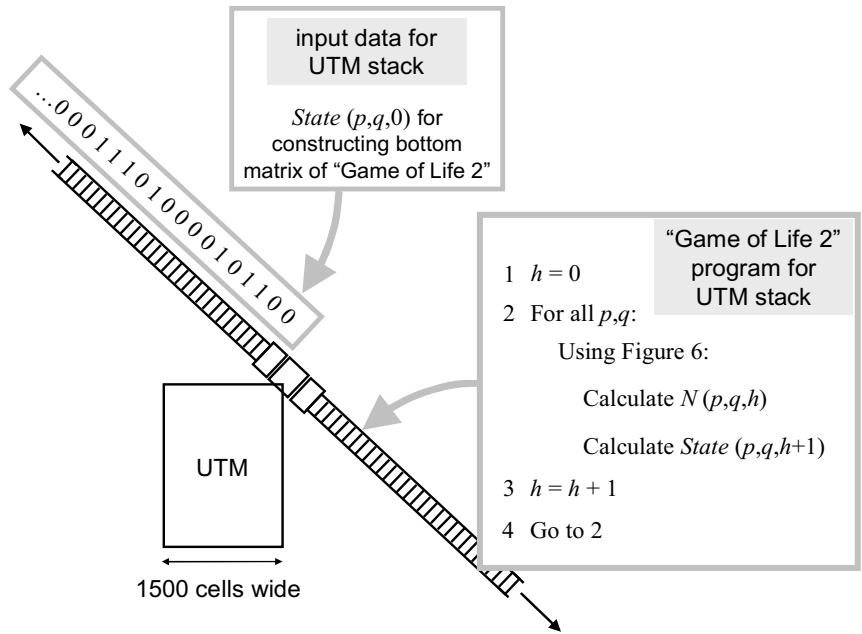

**Figure 7**: The UTM in the Game of Life is approximately 1,500 cells wide and is serviced by two stacks, one containing a program and the other containing data for the program to process.

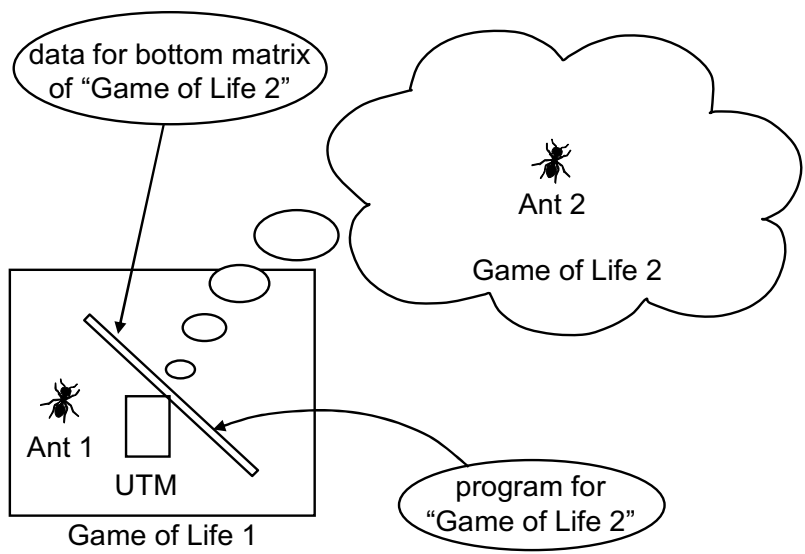

**Figure 8**: The Game of Life 2 is shown as a thought-bubble because it is not actually displayed on a computer screen in the way that Game of Life 1 is.

When Ant-1 builds the model of its environment, this can include features of itself (although, of course, the model will be incomplete for Gödel/Turing-type reasons). To that extent, Ant-1 may be said to be self-aware. By the same token, since Ant-2 is programmed similarly to Ant-1, Ant-2 may also be said to be self-aware. In both cases, this self-awareness extends to each knowing that it has a computer within itself – the Ants are aware that they compute. And so, Descartes-style, both Ants conclude that they exist!

Ant-1 will be aware that there is an Ant-2 embedded within the program and input data contained in UTM's stacks. Given a little licence, we can even imagine Ant-1 being smug about the fact that Ant-2 is not displayed on a computer screen, and so Ant-1 thinks that the existence of Ant-2 is in some sense less valid than its own, because Ant-2 is just a mathematical structure. The irony, of course, is that Ant-1 is, itself, "just" a mathematical structure – as we noted above, the computer program running Game of Life 1 is isomorphic to a static mathematical structure.

Of course, the mathematical structure of our own universe and multiverse is more complex than that of the Game of Life, but we share with Ant-1 and Ant-2 the essence of our self-awareness and the conviction that we exist.

### 9. Why do we exist?

Why does the Plexus contain such an apparently unlikely set of axioms that combine to produce our own distinctive multiverse? The answer is that the Plexus contains *every possible* mathematical structure. Most of these are sterile. Structures include cubes and spheres, sets like Mandelbrot's, and weird and splendid universes and multiverses, themselves containing beautiful substructures like stars and galaxies, exotic crystals and spectacular arrays of colour. In at least one of these multiverses, substructures exist that can gaze upon their universe and be moved by its grandeur.

Does the Plexus hypothesis make falsifiable predictions? Fundamentally, it is the framework – the structure – that needs to be tested, rather than quantum mechanics, of which the Plexus hypothesis is essentially an interpretation and with which it therefore ought to be compatible. The one difference is that, since the number of parallel block universes in our multiverse is not infinite, quantum probabilities should be rational rather than real numbers, as noted above. However, the hypothesis does not give a limit to the number of parallel universes, which may therefore be too great for experiment to distinguish between the two types of probability.

Tegmark argues [20] that a key prediction of a mathematical universe hypothesis is that physics in the future will continue to uncover mathematical regularities in nature. No doubt this prediction will turn out to be right, but it is difficult to see just what kind of experiment could falsify it. If, for example, it appears from a series of experiments that a physical system has no mathematical regularity (say, the weather), then, far from than falsifying the hypothesis, that very discovery will just motivate us to pursue the matter in the hope of uncovering regularity at a deeper level (say, the rules of chaos). The discovery of such an irregular system would not be cause to abandon the Plexus hypothesis.

So, if the Plexus hypothesis has not yet yielded testable predictions apart from those already intrinsic to quantum mechanics (with the difference between rational and real probabilities being too small to detect), what, then, is its attraction? The appeal of the Plexus hypothesis lies in its power to reconcile and explain two extraordinary observations about our universe: (1) part of your future is already my past, if I am moving towards you and (2) quantum entanglement experiments show that the universe is non-local.

The Many-Worlds interpretation of quantum mechanics comes close to accounting for the second observation, but it is nevertheless ruled out on several counts, the most important of which is the first of the above two observations. This observation means that the outcome of any quantum experiment in the universe is unique in that universe's future, which excludes the branching topology of the Many-Worlds picture. This was the motivation for postulating a multiverse of complete, stand-alone, parallel block universes.

In addition to this reconciliation of the two observations, explanations for other phenomena now emerge "for free". In particular, the real origin of the "arrow of time", conventionally associated with increasing entropy in our universe, arises from considering information across the whole multiverse. An individual universe (one without a gravitational force) could even be at a moment of maximal entropy, so that quantum fluctuations subsequent to that moment would not increase the entropy, with no consequent direction for time. However, while the random pattern at the moment of maximal entropy is initially repeated identically across the whole group of parallel universes with the same history, the pattern is generally different in each of the parallel universes at a later time, which therefore takes more information to describe. So it is the multiverse view that gives the arrow of time its direction. This same perspective also explains why time-symmetric quantum formulae are nevertheless consistent with time having a direction.

The Plexus hypothesis raises more questions than it answers. How might the equations of quantum and relativity be recast in terms of natural numbers as required by the hypothesis? What might be the quantum wave function at the end of time (as time is necessarily finite for the same reason that the number of parallel, block universes is finite)? Is the phenomenon of self-awareness a feature of only the lowest levels of the Plexus (individual parallel universes) or might consciousness exist at multiverse levels or even higher?

In contrast to the soothsayers who arise every generation to prognosticate the end of physics, we are now aware of more questions to be answered about our universe than ever before in our history. The Plexus offers up a trove of new horizons to explore in a quest the end of which humankind may well never reach.